\documentclass[aps,nofootinbib,floatfix]{revtex4}

\usepackage{amssymb}
\usepackage{amsmath}
\usepackage{graphicx}

\begin{document}

\title{Time Variable Cosmological Constants from Cosmological Horizons}

\author{Lixin Xu\footnote{Corresponding author}}
\email{lxxu@dlut.edu.cn}
\affiliation{Institute of Theoretical
Physics, School of Physics \& Optoelectronic Technology, Dalian
University of Technology, Dalian, 116024, P. R. China}
\author{Jianbo Lu}
\affiliation{Department of Physics, Liaoning Normal University,
Dalian 116029, P.R. China}
\author{Wenbo Li}
\affiliation{Institute of Theoretical Physics, School of Physics \&
Optoelectronic Technology, Dalian University of Technology, Dalian,
116024, P. R. China}

\begin{abstract}
In this paper, motivated from the fact that a de Sitter cosmological
boundary corresponds to a positive cosmological constant, we
consider time variable cosmological constants, dubbed {\it horizon
cosmological constants}. The horizon cosmological constants
correspond to Hubble horizon, future event horizon and particle
horizon are discussed respectively. When the Hubble horizon is taken
as a cosmological length scale, the effective equation of state of
horizon cosmological constant is quintessence-like. The values of
model parameter $c$ will determine the current status of our
universe. When particle horizon is taken as the cosmological length
scale, non viable cosmological model can be obtained for the
requirement of $\Omega_{\Lambda}<1/3$ which conflicts with current
comic observations. When the future event horizon is taken as the
role of cosmological length scale, the forms of effective equation
of state of horizon cosmological constants are the the same as the
holographic ones. But, their evolutions are different because of the
effective interaction with cold dark matter.
\end{abstract}



\maketitle

\section{Introduction}

The observation of the Supernovae of type Ia
\cite{ref:Riess98,ref:Perlmuter99} provides the evidence that the
universe is undergoing an accelerated expansion. Combining the
observations from Cosmic Background Radiation
\cite{ref:Spergel03,ref:Spergel06} and SDSS
\cite{ref:Tegmark1,ref:Tegmark2}, one concludes that the universe at
present is dominated by $70\%$ exotic component, dubbed dark energy,
which has negative pressure and pushes the universe to accelerated
expansion. Of course, a natural explanation to the accelerated
expansion is due to a positive tiny cosmological constant. Though,
it suffers the so-called {\it fine tuning} and {\it cosmic
coincidence} problems. However, in $2\sigma$ confidence level, it
fits the observations very well \cite{ref:Komatsu}. If the
cosmological constant is not a real constant but is time variable,
the fine tuning and cosmic coincidence problems can be removed. In
fact, this possibility was considered in the past years.

In particular, the dynamic vacuum energy density based on
holographic principle are investigated extensively
\cite{ref:holo1,ref:holo2}. According to the holographic principle,
the number of degrees of freedom in a bounded system should be
finite and has relations with the area of its boundary. By applying
the principle to cosmology, one can obtain the upper bound of the
entropy contained in the universe. For a system with size $L$ and UV
cut-off $\Lambda$ without decaying into a black hole, it is required
that the total energy in a region of size $L$ should not exceed the
mass of a black hole of the same size, thus $L^3\rho_{\Lambda} \le L
M^2_{P}$. The largest $L$ allowed is the one saturating this
inequality, thus $\rho_{\Lambda} =3c^2 M^{2}_{P} L^{-2}$, where $c$
is a numerical constant and $M_{P}$ is the reduced Planck Mass
$M^{-2}_{P}=8 \pi G$. It just means a {\it duality} between UV
cut-off and IR cut-off. The UV cut-off is related to the vacuum
energy, and IR cut-off is related to the large scale of the
universe, for example Hubble horizon, future event horizon or
particle horizon as discussed by
\cite{ref:holo1,ref:holo2,ref:Horvat1,ref:Horvat2}. The holographic
dark energy in Brans-Dicke theory is also studied in Ref.
\cite{ref:BransDicke,ref:BDH1,ref:BDH2,ref:BDH3,ref:BDH4,ref:BDH5}.

In the standard and Brans-Dicke holographic dark energy models when
the Hubble horizon is taken as the role of IR cut-off,
non-accelerated expansion universe can be achieved
\cite{ref:holo1,ref:holo2,ref:BransDicke}. However, the Hubble
horizon is the most natural cosmological length scale, how to
realize an accelerated expansion by using it as an IR cut-off will
be interesting. In fact, an accelerated expansion universe has been
realized in the holographic case when the model parameter $c$ is
time variable \cite{ref:HoloXu}.

As known, for any nonzero value of the cosmological constant
$\Lambda$, a natural length scale and time scale
\begin{equation}
r_{\Lambda}=t_{\Lambda}=\sqrt{3/|\Lambda|}
\end{equation}
can be introduced into the Einstein's theory. Reversely, a
cosmological length scale and time scale may introduce a
cosmological constant or vacuum energy density into Einstein's
theory. Of course, the important is how to choose a proper
cosmological length scale or time scale to obtain a tiny
cosmological constant or vacuum energy density. Honestly, we have
not the first physical principle to determine the length or time
scale. But, one can immediately relate these length or time scales
to the biggest natural scales, for example Hubble horizon, particle
horizon and future event horizon etc. Because the under-researched
system is the universe, for a cosmological length scale
$r_{\Lambda}(t)$, one has
\begin{equation}
\Lambda(t)=\frac{3}{r^2_{\Lambda}(t)}.
\end{equation}
When the length scale is time variable, a time variable cosmological
constant can be obtained. Inspired by this observation, one can
consider time variable cosmological constant from this analogue and
let alone the holographic principle. The important is that with this
analogue an accelerated expansion will be obtained when the Hubble
horizon is taken as a cosmological length scale. If the other
cosmological length scales, for examples future event horizon and
particle horizon, are put in, a time variable tiny cosmological
constant and an accelerated expansion universe will be obtained. In
this paper, we will explore these possibilities. For their explicit
relations with the cosmological length scales, {\it i.e.}
cosmological horizons, we dub them {\it horizon cosmological
constants} as contrasts to the {\it holographic cosmological
constants} \cite{ref:Horvat1,ref:Horvat2,ref:Feng}. For time scale
case, a time variable cosmological constant has been discussed in
\cite{ref:xuCCA}.

This paper is structured as follows. In Section \ref{sec:TVC}, we
give a brief review of time variable cosmological constant. In
Section \ref{sec:HHL}, \ref{sec:FEH} and \ref{sec:PHL}, Hubble
horizon, future event horizon and particle horizon will be
considered as cosmological length scale respectively and their
corresponding time variable cosmological constants are discussed. We
will discuss their evolutions in Section \ref{sec:evolution}.
Conclusions are set in Section \ref{sec:Con}.

\section{Time Variable Cosmological Constant}\label{sec:TVC}

The Einstein's equations with a cosmological constant can be written
as
\begin{equation}
R_{\mu\nu}-\frac{1}{2}Rg_{\mu\nu}+\Lambda g_{\mu\nu}=8\pi G
T_{\mu\nu},\label{eq:EE}
\end{equation}
where $T_{\mu\nu}$ is the energy-momentum tensor of ordinary matter
and radiation. This equation can be obtained via the variation of
the cosmological constant contained Einstein-Hilbert action with
respect to $g^{\mu\nu}$, i.e. $\delta S_4/\delta g^{\mu\nu}=0$. And
the result does not depend on the cosmological constant is time and
space variable or not. From the Bianchi identity, one has the
conservation of the energy-momentum tensor
$\nabla^{\mu}T_{\mu\nu}=0$, it follows necessarily that $\Lambda$ is
a fixed constant. If the cosmological constant permeates smoothly
the whole universe and is only time variable, say
$\Lambda=\Lambda(t)$, one can move the cosmological constant to the
right hand side of Eq. (\ref{eq:EE}) and takes
$\tilde{T}_{\mu\nu}=T_{\mu\nu}-\frac{\Lambda(t)}{8\pi G}g_{\mu\nu}$
as the total energy-momentum tensor. Once again to preserve the
Bianchi identity or local energy-momentum conservation law,
$\nabla^{\mu}\tilde{T}_{\mu\nu}=0$, one has, in a spacially flat FRW
universe,
\begin{equation}
\dot{\rho}_{\Lambda}+\dot{\rho}_{m}+3H\left(1+w_{m}\right)\rho_{m}=0,\label{eq:conservation}
\end{equation}
where $\rho_{\Lambda}=M^2_{P}\Lambda(t)$ is the energy density of
time variable cosmological constant and its equation of state is
$w_{\Lambda}=-1$, and $w_{m}$ is the equation of state of ordinary
matter, for the cold dark matter $w_m=0$. It is natural to consider
interactions between variable cosmological constant and dark matter
\cite{ref:Horvat2}, as seen from Eq. (\ref{eq:conservation}). After
introducing an interaction term $Q$, one has
\begin{eqnarray}
\dot{\rho}_{m}+3H\left(1+w_{m}\right)\rho_{m}=Q,\label{eq:rhom} \\
\dot{\rho}_{\Lambda}+3H\left(\rho_{\Lambda}+p_{\Lambda}\right)=-Q,\label{eq:rholambda}
\end{eqnarray}
and the total energy-momentum conservation equation
\begin{equation}
\dot{\rho}_{tot}+3H\left(\rho_{tot}+p_{tot}\right)=0.
\end{equation}
For a time variable cosmological constant, the equality
$\rho_{\Lambda}+p_{\Lambda}=0$ still holds. Immediately, one has the
interaction term $Q=-\dot{\rho}_{\Lambda}$ which is different from
the interactions between dark matter and dark energy considered in
the literatures \cite{ref:interaction} where a general interacting
form $Q=3b^2H\left(\rho_{m}+\rho_{\Lambda}\right)$ is put by hand.
With observation to Eq. (\ref{eq:rholambda}), the interaction term
$Q$ can be moved to the left hand side of the equation, and one has
the effective pressure of the time variable cosmological
constant-dark energy
\begin{equation}
\dot{\rho}_{\Lambda}+3H\left(\rho_{\Lambda}+p^{eff}_{\Lambda}\right)=0,
\end{equation}
where $p^{eff}_{\Lambda}=p_{\Lambda}+\frac{Q}{3H}$ is the effective
dark energy pressure. Also, one can define the effective equation of
state of dark energy
\begin{eqnarray}
w^{eff}_{\Lambda}&=&\frac{p^{eff}_{\Lambda}}{\rho_{\Lambda}}\nonumber\\
&=&-1+\frac{Q}{3H\rho_{\Lambda}}\nonumber\\
&=&=-1-\frac{1}{3}\frac{d \ln \rho_{\Lambda}}{d\ln
a}.\label{eq:EEOS}
\end{eqnarray}
Obviously, when $\Lambda$ is a fixed constant, the effective
interaction term $Q$ becomes zero. Then the $\Lambda$CDM universe is
recovered. The Friedmann equation as usual can be written as, in a
spacially flat FRW universe,
\begin{equation}
H^2=\frac{1}{3M^2_P}\left(\rho_{m}+\rho_{\Lambda}\right)\label{eq:FE}.
\end{equation}

\section{Horizon cosmological constants}

\subsection{Hubble horizon as cosmological length scale}\label{sec:HHL}

In fact, Horvat has considered this possibility from holographic
principle \cite{ref:Horvat2}, where the Hubble horizon $H^{-1}$ was
taken as a cosmological length scale. When Hubble horizon $H^{-1}$
is chosen, one obtains a time variable cosmological constant
\begin{equation}
\Lambda(t)=3c^2H^{2}(t)
\end{equation}
which is just the one considered by Horvat \cite{ref:Horvat2}, where
$c$ is a constant. As known, our universe is filled with dark matter
and dark energy and deviates from a de Sitter one. Just to describe
this gap, the constant $c$ is introduced. It can be seen that a
$c<1$ constant is expected under the consideration of the energy
budget of our universe and the de Sitter universe will be recovered
when $c=1$. Now, the corresponding vacuum energy density can be
written as
\begin{equation}
\rho_{\Lambda}=3c^2M^2_{P}H^2
\end{equation}
which has the same form as the so-called holographic dark energy
which is based on holographic principle. With this vacuum energy,
the Friedmann equation (\ref{eq:FE}) can be rewritten as
\begin{equation}
\rho_{m}=3(1-c^2)M^2_PH^2.
\end{equation}
To protect a positive dark matter energy density, a constraint
\begin{equation}
c^2<1
\end{equation}
is required. Immediately, a scaling solution is
obtained
\begin{equation}
\frac{\rho_{m}}{\rho_{\Lambda}}=\frac{1-c^2}{c^2}\label{eq:scaling}.
\end{equation}
Substituting Eq. (\ref{eq:scaling}) into Eq.
(\ref{eq:conservation}), one has
\begin{equation}
\rho_{\Lambda}=\frac{c^2}{1-c^2}\rho_{m}\sim a^{-3(1-c^2)}.
\end{equation}
Here, one can see a rather different result on $\rho_{m}$ from the
standard evolution $a^{-3}$. In this case, the deceleration
parameter becomes
\begin{eqnarray}
q&=&-\frac{\ddot{a}a}{\dot{a}^2}=-\frac{\dot{H}+H^2}{H^2}\nonumber\\
&=&\frac{1}{2}-\frac{3}{2}c^2.
\end{eqnarray}
To obtain a current accelerated expansion universe, i.e. $q<0$, and
to protect positivity of dark matter energy density, one obtains a
constraint to the constant $c$
\begin{equation}
1/3<c^2<1.\label{eq:cconstraint}
\end{equation}
The effective equation of state of vacuum energy density is
\begin{eqnarray}
w^{eff}_{\Lambda}&=&-1-\frac{1}{3}\frac{d \ln \rho_{\Lambda}}{d\ln
a}\nonumber\\
&=&-c^2.
\end{eqnarray}
Under the constraint Eq.(\ref{eq:cconstraint}), one can see that a
quintessence like dark energy is obtained. This is tremendous
different from holographic dark energy model where non-accelerated
expansion universe can be achieved when the Hubble horizon taken as
the IR cut-off \cite{ref:holo1,ref:holo2,ref:BransDicke}. Also, it
is easily that the de Sitter universe will be recovered when $c=1$.
Once the constant $c$ deviates from $c=1$, a scaling solution will
be obtained. It is clear that the properties of horizon cosmological
constant depend on the value of the model parameter $c$. But, it can
not describe the whole evolution of our universe for the lack of
transition from decelerated expansion to accelerated expansion. With
a fixed constant $c$, the deceleration parameter $q$ is also a fixed
constant. So, it will be not a variable dark energy model when the
Hubble horizon as a cosmological length scale. However, it can be
remedied by treating $c$ a no fixed constant \cite{ref:HoloXu}. So,
other candidates to the cosmological length scale would be chosen.

\subsection{Future event horizon as a cosmological length scale}\label{sec:FEH}

The future event horizon is defined as
\begin{equation}
R_{e}=a\int^{\infty}_{t}\frac{dt'}{a}=a\int^{\infty}_{a}\frac{da'}{Ha'^{2}}\label{eq:eh}
\end{equation}
which is the boundary of the volume a fixed observer may eventually
observe. Taking it as the role of cosmological length scale, one has
the vacuum energy density
\begin{equation}
\rho_{\Lambda}=3c^2M^2_P/R^2_e.\label{eq:ehvv}
\end{equation}
Via the definition of the dimensionless energy densities
$\Omega_{m}=\rho_{m}/(3M^2_PH^2)$ and
$\Omega_{\Lambda}=\rho_{\Lambda}/(3M^2_PH^2)$, the Friedmann
equation is rewritten as
\begin{equation}
\Omega_{m}+\Omega_{\Lambda}=1.
\end{equation}
The energy conservation equation (\ref{eq:conservation}) becomes
\begin{equation}
\frac{d\ln
H}{dx}+\frac{3}{2}\left(1-\Omega_{\Lambda}\right)=0,\label{eq:diffH}
\end{equation}
where $x=\ln a$. Combining Eq. (\ref{eq:eh}), Eq. (\ref{eq:ehvv})
and the definition of the dimensionless energy density
$\Omega_{\Lambda}$, one has
\begin{equation}
\int^{\infty}_{a}\frac{d\ln
a'}{Ha'}=\frac{c}{aH}\sqrt{\frac{1}{\Omega_{\Lambda}}}.\label{eq:integral}
\end{equation}
Taking the derivative with respect to $x=\ln a$ from the both sides
of the above equation (\ref{eq:integral}), one has the differential
equation
\begin{equation}
\frac{d\ln
H}{dx}+\frac{1}{2}\frac{d\ln\Omega_{\Lambda}}{dx}=\frac{\sqrt{\Omega_{\Lambda}}}{c}-1.\label{eq:Hdx}
\end{equation}
Substituting Eq. (\ref{eq:diffH}) into above differential equation
(\ref{eq:Hdx}), one obtains the differential equation of
$\Omega_{\Lambda}$
\begin{equation}
\Omega_{\Lambda}'=\Omega_{\Lambda}\left(1-3\Omega_{\Lambda}+\frac{2}{c}\sqrt{\Omega_{\Lambda}}\right),\label{eq:EHEQ}
\end{equation}
where $'$ denotes the derivative with respect to $x=\ln a$. This
equation describes the evolution of the dimensionless energy density
of dark energy. Clearly, one can see that this equation is different
from the corresponding one derived in \cite{ref:holo2} for its
different form. Given the initial conditions $\Omega_{\Lambda0}$,
the evolution of the horizontal cosmological constant will be
obtained.
Once the values of $c$ and $\Omega_{\Lambda0}$ are fixed,
$\Omega_{\Lambda}$, as a function of $\ln a$ or redshift $z$,
describes the evolution of the dark energy completely. In fact, the
model parameter $c$ can be determined by the cosmic observations
such as SN Ia, BAO and CMB shift parameter $R$. From Eq.
(\ref{eq:EEOS}), it is easy to obtain the effective equation of
state of dark energy
\begin{eqnarray}
w^{eff}_{\Lambda}&=&-1-\frac{1}{3}\frac{d \ln \rho_{\Lambda}}{d\ln
a}\nonumber\\
&=&-\frac{1}{3}\left(1+\frac{2}{c}\sqrt{\Omega_{\Lambda}}\right).
\end{eqnarray}
It is in the range of $-(1+\frac{2}{c})/3<w^{eff}_{\Lambda}<-1/3$,
when one notices the dark energy density ratio
$0\le\Omega_{\Lambda}\le1$. The form of the effective equation of
state of the horizon cosmological constant is the same as the one of
holographic dark energy. However, for the differences of the
evolution of $\Omega_{\Lambda}$, its behaviors will be different
from the holographic one. The deceleration parameter can be easily
derived
\begin{eqnarray}
q&=&-\frac{\dot{H}+H^2}{H^2}\nonumber\\
&=&-1-\frac{d\ln H}{d\ln a}\nonumber\\
&=&\frac{1}{2}-\frac{3}{2}\Omega_{\Lambda}.
\end{eqnarray}
To have an current accelerated expansion of the universe,
$\Omega_{\Lambda0}>1/3$ is required.

\subsection{Particle horizon as cosmological length
scale}\label{sec:PHL}

In this part, we will consider another possible candidate to the
cosmological length scale-the particle horizon which is defined as
\begin{equation}
R_{p}=a(t)\int^{t}_{0}\frac{dt'}{a}=a\int^{a}_{0}\frac{da'}{Ha'^{2}}.
\end{equation}
It is the length scale that a particle can pass from the beginning
of the universe. In this case, the vacuum energy density is given as
\begin{equation}
\rho_{\Lambda}=3c^2M^2_P/R^2_p.\label{eq:ehv}
\end{equation}
Repeating the analysis and calculations as done in \ref{sec:FEH},
one has the differential equation of $\Omega_{\Lambda}$
\begin{equation}
\Omega_{\Lambda}'=\Omega_{\Lambda}\left(1-3\Omega_{\Lambda}-\frac{2}{c}\sqrt{\Omega_{\Lambda}}\right),\label{eq:PHEQ}
\end{equation}
where $'$ denotes the derivative with respect to $x=\ln a$.
The effective equation of state and deceleration parameter, which
are the same as that of the future event horizon case, are given as
\begin{eqnarray}
w^{eff}_{\Lambda}&=&-\frac{1}{3}\left(1-\frac{2}{c}\sqrt{\Omega_{\Lambda}}\right),\\
q&=&\frac{1}{2}-\frac{3}{2}\Omega_{\Lambda}.
\end{eqnarray}
In terms of redshift $z$, the evolution equation (\ref{eq:PHEQ}) can
be recast into
\begin{equation}
\frac{d\Omega_{\Lambda}}{dz}=-\frac{1}{(1+z)}\Omega_{\Lambda}\left(1-3\Omega_{\Lambda}-\frac{2}{c}\sqrt{\Omega_{\Lambda}}\right).
\end{equation}
As known, the dimensionless parameter of $\Omega_{\Lambda}(z)$
decreases with the redshift $z$. So, the required condition will be
$\frac{d\Omega_{\Lambda}}{dz}<0$. Then one has the constraint to the
expression $(1-3\Omega_{\Lambda}-2\sqrt{\Omega_{\Lambda}}/c)>0$. By
simple calculation, one has
\begin{equation}
0<\frac{2}{c}<\frac{1}{\sqrt{\Omega_{\Lambda}}}-3\sqrt{\Omega_{\Lambda}},
\end{equation}
which is a strong constraint and requires $\Omega_{\Lambda}<1/3$
always. Comparing the result that current value of
$\Omega_{\Lambda}$ is $\sim0.7$, one can conclude that particle
horizon case can not give a viable cosmological model.

\subsection{Evolution curves of horizon cosmological
constants}\label{sec:evolution}

\begin{figure}[tbh]
\centering
\includegraphics[width=3in]{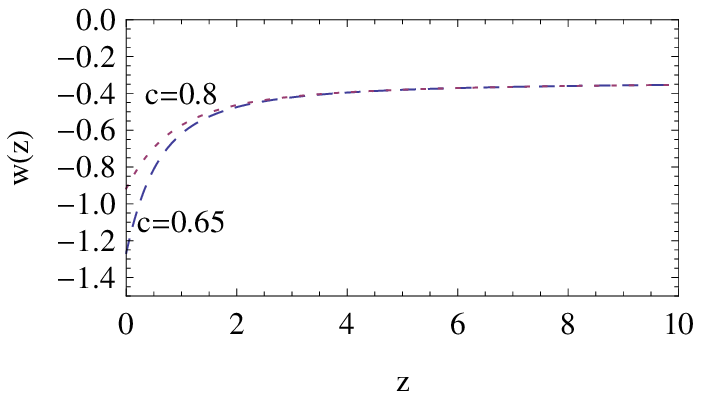}
\includegraphics[width=3in]{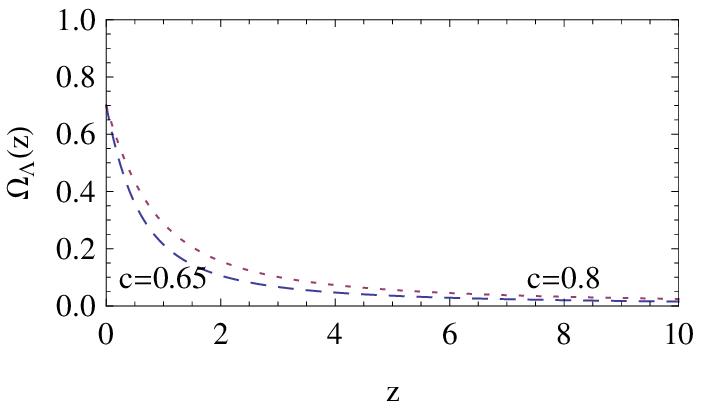}
\caption{The evolutions of effective equation of state
$w_{\Lambda}(z)$ (left panel) and dimensionless density parameter
$\Omega_{\Lambda}(z)$ (right panel) of horizon cosmological
constants with respect to the redshift $z$, where the values
$\Omega_{\Lambda0}=0.70$, $c=0.65$ (dashed lines), $c=0.8$ (dotted
lines) are adopted.}\label{fig:evolution}
\end{figure}

\begin{figure}[tbh]
\centering
\includegraphics[width=3in]{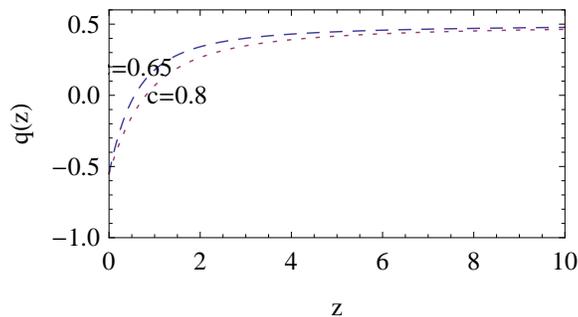}
\caption{The evolutions of deceleration parameter $q(z)$ with
respect to the redshift $z$, where the values
$\Omega_{\Lambda0}=0.70$, $c=0.65$ (dashed lines), $c=0.8$ (dotted
lines) are adopted.}\label{fig:qzevolution}
\end{figure}

Now, it is proper to discuss the evolutions of horizon cosmological
constants. Here, we just discuss the evolutions of effective
equation of state and dimensionless energy density. In the
holographic dark energy model, the value of the parameter $c$
determines the property of holographic dark energy. When $c>1$,
$c=1$ and $c<1$ in the case of the event horizon as an IR cut-off,
the holographic dark energy behaves like quintessence, cosmological
constant and phantom respectively. In our cases, there are some
differences and common grounds. In the Hubble horizon cosmological
constant case, the effective equation of state of time variable
cosmological constant is always quintessence like or true constant
when the value of $c$ is $c<1$ and $c=1$ respectively. However, in
the corresponding holographic case, non-accelerated expansion
universe can be obtained when Hubble horizon as an IR cut-off. In
the future event horizon case, the behaviors of effective equation
of state determined by $c$ are the same as that of the holographic
ones, however the evolutions are different. We have plotted the
evolutions of the effective equation of state and dimensionless
energy density with respect to the redshift $z$ in Fig.
\ref{fig:evolution} where different values of model parameter $c$
are adopted.
Clearly by setting $q=0$, one can obtain the transition redshift
$z_T$ form decelerated expansion to accelerated expansion which is
the solution of $\Omega_{\Lambda}(z)=1/3$. For example, let
$\Omega_{\Lambda0}=0.70$ and $c=0.65$, one has the transition
redshift $z_T\approx0.56$. In fact, the model parameters $c$ and
$\Omega_{\Lambda0}$ can be fixed by fitting current cosmic
observations \cite{ref:ChenZhuXU}. The evolutions of deceleration
parameter are shown in Fig. \ref{fig:qzevolution}.

\section{Conclusions}\label{sec:Con}

In this paper, time variable cosmological constants, dubbed {\it
horizon cosmological constants}, as analogues to the fact that a de
Sitter cosmological boundary corresponds to a positive cosmological
constant, are explored. The horizon cosmological constants
correspond to Hubble horizon, future event horizon and particle
horizon are discussed respectively. When the Hubble horizon is taken
as a cosmological length scale, the effective equation of state of
horizon cosmological constant $w^{eff}_{\Lambda}=-c^2$ is
quintessence like. In this case, the deceleration parameter
$q=1/2-3c^2/2$ is negative when $1/3<c^2<1$, i.e. an accelerated
expansion universe is obtained. That is tremendous different from
the holographic dark energy model in Einstein' gravity theory and
Brans-Dicke theory where non-accelerated expansion universe can
exist when the Hubble horizon as an IR cut-off. It also can be seen
that, once the constant $c=1$, the de Sitter universe will be
recovered. However, with a fixed constant $c$, the deceleration
parameter $q$ is also a fixed constant. So, it will be not a
variable dark energy model when the Hubble horizon as a cosmological
length scale for the lack of transition from decelerated expansion
to accelerated expansion. But, it can be remedied by treating $c$ a
no fixed constant. So, other candidates to the cosmological length
scale would be chosen and checked. When particle horizon is taken as
the cosmological length scale, non viable cosmological model can be
obtained for the requirement of $\Omega_{\Lambda}<1/3$ which
conflicts with current comic observations. When future event horizon
is taken as the cosmological length scale, the forms of effective
equation of state of horizon cosmological constants are the same as
the holographic ones. But, their evolutions and behaviors are
different from the holographic one for their differences between the
evolution equations because of the effective interaction with cold
dark matter, see Eq.(\ref{eq:EHEQ}) and Eq.(\ref{eq:PHEQ}). These
also can be seen from the evolution curves in Fig.
\ref{fig:evolution}.

\acknowledgements{This work is supported by NSF (10703001), SRFDP
(20070141034) of P.R. China and the Fundamental Research Funds for
the Central Universities (DUT10LK31).}

\end{document}